\documentstyle{article}

\begin{document}

\title{{\bf Exact Solution of an Evolutionary Model without Ageing}} 
\author{Roberto N. Onody \thanks{onody@ifsc.sc.usp.br} and
Nazareno G. F. de Medeiros\thanks{ngfm@ifsc.sc.usp.br}\\ \\ 
{\small {\em Departamento de F\'{\i}sica e Inform\'{a}tica } }\\ 
{\small {\em Instituto de F\'{\i}sica de S\~{a}o Carlos} }\\ 
{\small {\em Universidade de S\~{a}o Paulo - Caixa Postal 369} }\\ 
{\small {\em 13560-970 - S\~{a}o Carlos, S\~{a}o Paulo, Brasil.}}}
\date{} 
\maketitle 
\normalsize 
\baselineskip=16pt

\begin{abstract} 

We introduce an age-structured asexual population model containing all the 
relevant features of evolutionary ageing theories. Beneficial as well as 
deleterious mutations, heredity and arbitrary fecundity are present and
managed by natural selection. An exact solution without ageing is found. We
show that fertility is associated with generalized forms of the Fibonacci 
sequence, while mutations and natural selection are merged into an integral 
equation which is solved by Fourier series. Average survival probabilities and 
Malthusian growth exponents are calculated indicating that the system may 
exhibit mutational meltdown. The relevance of the model in the context of
fissile reproduction groups as many protozoa and coelenterates
is discussed.

\vspace{2cm}


\end{abstract}

\newpage

\section{Introduction}

\indent

For all alive organisms death is inexorable. If life is not abbreviated by
disease, predation or accidents then death is preceded by senescence \cite{char,
finch}. Senescence seems to be related to reproductive strategies and it starts 
after an individual reaches the fertile age. It can be a slow process for 
species which reproduce many times (iteroparous) or an abrupt process for 
species reproducing only once (semelparous) - the catastrophic senescence of the
Pacific salmon being a good example of the late.  

There are tree kinds of ageing theories: biochemical, evolutionary 
\cite{rose,wa} and telomeric \cite{re}. 
The biochemical invokes damages on DNA, cells, tissues and organs. Defective
proteins are synthesized altering the normal course of metabolism. The presence
of free radicals, i. e., of unpaired highly reactive electrons, can cause death
of the cells or may even lead to cancer. As a consequence, such theories are 
today firmly connected to modern gerontology. Biochemical theories predict that
species with higher metabolism would have shorter lifetime. A criticism against 
this result arises when the lifetime of birds and mammals (of the same size
and under optimal conditions in captivity) are compared. Usually, birds live 
longer. Even closely related organisms like bats and rodents with comparable 
sizes have very different lifetimes. Such differences should be explained by the
evolutionary theories of ageing. In the telomere hypothesis of senescence, 
replication of normal cells is accompanied by a telomeric shortening. This 
acts as a mitotic clock resulting in a permanent exit of the cell cycle.

Evolutionary theories of ageing are hypothetico-deductive in character,
not inductive. They fall into two classes: the optimality 
theory and the mutational theory. In the optimality
theory \cite{par,pb}, fitness is maximized by increasing the
survival and reproduction rate early in life at the expense of late, i. e., 
it sees senescence as a necessary cost of processes beneficial to youth.
On the other hand, the mutational theory \cite{med} explains senescence 
as a result of late-acting deleterious mutations, i. e., that a greater 
mutation load on the last part of the life is less strongly selected.

In this paper, we obtain both analytical and numerical solutions for a simple 
age-structured population model. In this model, reproduction is asexual 
and the individuals are submitted to helpful or deleterious hereditary 
mutations. The number of offsprings is fixed but arbitrary otherwise and 
the population dynamics
is managed by natural selection. Amazingly, this system does not exhibit 
senescence even when deleterious mutations are predominant. From a mathematical
point of view, the solution we found factorizes into a fertility and a 
mutational sectors. It is shown that the fertility sector is completely 
described by generalized forms of Fibonacci sequences. On the other hand, 
the mutational sector, solved by Fourier series, incorporates the combined 
effects of mutation and natural selection. If harmful mutation is intense then 
a mutational meltdown \cite{ly} process can take place. All these results were 
corroborated by Monte Carlo simulations.
From a biological point of view, 
our results make the model a good candidate to describe groups in which all 
reproduction occurs by fission, such as protozoa and a 
miscellany of coelenterates, since all them appear to lack ageing \cite{rose}.
 
\section{The Model}

\indent

Consider a population distributed by $L+1$ ages $i$ ($i=0,1,...,L$) with 
respective birth rates $m_{i}$. 
We call babies the individuals at age $0$. They do not reproduce, i. e.,
$m_{0}=0$. Individuals with age $i=L$ die after reproduction. Let $N_{i}(J,t)$ 
be the number of individuals at age $i$ with 
survival probability $J$ between $J$ and $J+dJ$ at time $t$ 
(of course, $J \in [0,1]$). We choose, as initial condition, an uniform 
distribution of babies, i. e., $N_{i}(J,0)=N_{0} \; \delta _{i,0} $ with 
$N_{0}$ being a constant. Let us point out that senescence here means that the 
average survival probability drops with age $i$. 
At time $t$, each individual is submitted to 
mutation which changes its survival probability from $J$ to $J'$ in the 
following way

\begin{equation}
J'=J \; e^{\epsilon}
\end{equation}
where $\epsilon$ is a random number chosen in the interval $[-a,b]$
($a > 0$ and $b > 0$), generated by the uniform probability density 
distribution $q(\epsilon) = \frac{H(\epsilon+a) - H(\epsilon -b)}{a+b}$ ($H(x)$ 
stands for the Heaviside function). If $a > b$ then deleterious mutations 
are dominant. The transition probability $P(J',J)$, that an individual has its 
survival probability changed from $J$ to $J'$ is given by the integral
$P(J',J)=\int_{-\infty}^{+\infty} \delta (J'-J e^{\epsilon})
q(\epsilon) d \epsilon $. Observing that a helpful mutation is allowed as long
as it does not increase the survival probability beyond unity, we get

\begin{eqnarray}
J' (a+b) P(J',J) & = 
\{ [H(J'- J e^{-a})- H(J'- J e^{b})][H(J) - H(J - e^{-b})] \nonumber \\
& + [H(J'-J e^{-a}) - H(J'-1)][H(J- e^{-b}) - H(J-1)] \}
\end{eqnarray}

After mutation, all population passes through natural selection. The survivors
then reproduce, generating babies with inherited characteristics, i. e.,

\begin{equation}
N_{0}(J,t)= \sum_{i=1}^{L} m_{i} \; N_{i}(J,t) , \;\;\; for \; t \ge 1
\end{equation}

At time step $t+1$ individuals get older or die. Taking into account mutations
and natural selection (in this order), their number is given by

\begin{equation}
N_{i+1}(J',t+1) = \int_{0}^{1} J' \; P(J',J) \; N_{i}(J,t) \; dJ \;\;
for \; i=0, \dots, L-1
\end{equation}

Clearly, $N_{i}(J,t < i)=0$ for 
$i \ne 0$.
For $ t \ge 1$, equations (3) and (4) can be rewritten in the matricial form

\begin{equation}
\vec{N}(J',t+1) = M \int_{0}^{1} f(J',J) \; \vec{N}(J,t) \; dJ
\end{equation}
where $f(J',J) = J' \; P(J',J) $ and $\vec{N} (J,t)$ and $M$ are the 
following vector and matrix 

\begin{center}
$  \vec{N} (J,t) = \left(
\begin{array}{c}
N_{1} (J,t) \\ N_{2} (J,t) \\ N_{3} (J,t) \\  \vdots \\ N_{L} (J,t)
\end{array} \right)
\mbox{,} \;\;
M = \left(
\begin{array}{cccccc}
m_{1} & m_{2} & m_{3} & \dots & m_{L-1} & m_{L} \\
1  & 0  & 0  & \dots & 0       & 0  \\
0  & 1  & 0  & \dots & 0       & 0  \\
\vdots &  \vdots & \vdots & \mbox{} & \vdots & \vdots \\
0  & 0  & 0  & \dots & 1       & 0 
\end{array} \right)
$
\end{center}

Matrix $M$ is a special form of the Leslie matrices \cite{les}.
Iterating equation (5) we obtain at time $t \ge 1$
\begin{center}
\begin{equation}
\left(
\begin{array}{c}
N_{1} (J_{t},t) \\ N_{2} (J_{t},t) \\ N_{3} (J_{t},t) \\  \vdots \\ 
N_{L} (J_{t},t)
\end{array} \right)
\mbox{$ \;\; = \; N_{0} \; F(J_{t},t) \; M^{(t-1)} $}
\left(
\begin{array}{c}
1 \\ 0 \\ 0 \\  \vdots \\ 0
\end{array} \right)
\end{equation}
\end{center}
where 

\begin{equation}
F(J_{t},t) = \int_{0}^{1} \dots \int_{0}^{1} \prod_{j=1}^{t}
f(J_{j},J_{j-1}) \; dJ_{j-1}
\end{equation}
and $M^{0}$ is the $L \; \times L$ identity matrix.

\section{Generalized Fibonacci Sequences}

\indent

Equation (6) shows that
the dynamics factorizes into two sectors: the fertility,
exclusively contained in the matrix $M$, and the mutational, enclosing both
mutations and natural selection processes and represented by the 
function $F$. Clearly, at time $t$, we need only to know the first column
elements of the $(t-1)th$ power of the fertility matrix. As we shall see,
these elements form a Fibonacci sequence. Let us call $A_{L}(t+1-i)$ the 
element of the $ith$ line and first column. It is a simple exercise to 
verify that if 
$L=2$ and $m_{1}=m_{2}=1$ then $A_{2}(t)$ can be calculated, at any time $t$,
through  the relation $A_{2}(t) = A_{2}(t-1) + A_{2}(t-2)$, with 
$A_{2}(1)=A_{2}(2)=1$, which is exactly the Fibonacci sequence!. The ratio
of two sucessives numbers, $ \lim _{t \rightarrow \infty} A_{L}(t+1)/A_{L}(t) = 
1.618 \dots $ , gives the golden section or the divine proportion as called by
Kepler. It is astonishing to find this ubiquitous sequence in so
disparate things like the cluster-cluster  aggregates \cite{sore}, the division 
of a line into extreme and mean ratio, the pentagram star worn by the 
Pythagoreans, the continued fractions, the aesthetic proportions of 
the Parthenon at Athens \cite{hun}, and now, here, in population dynamics. 

When the number of ages is 4 ($L=3$) and $m_{1}=m_{2}=m_{3}=1$ then 
$A_{3}(t) = A_{3}(t-1) + A_{3}(t-2) + 
A_{3}(t-3)$, with $A_{3}(1)=A_{3}(2)=1$ and $A_{3}(3)=2$ (this sequence 
generates the so called tribonacci numbers). For an arbitrary number of ages
and fecundity we have

\begin{equation}
A_{L}(t) =\sum_{k=1}^{L} m_{k} \; A_{L}(t-k) \; , \;\; \mbox{for $t \ge (L+1)$)}
\end{equation}

The first $L$ numbers (necessary to initialize the sequence above) are 
evaluated through the 
identification $A_{L}(t) \equiv A_{2}(t)$, for $ t=1, \dots , L$. 
The numbers $A_{2}(t)$ are determined, on the other hand,  by using 
$A_{2}(1)=1$ and $A_{2}(2)=m_{1}$ in the expression above. 
Recurrence formulas like equation (8) are {\em generalized forms} of the 
Fibonacci sequence. These results, together with equation (6), allow us to
write down the number of individuals at time $t$ with age $i \ge 1$ and 
survival probabibity $J$ (we renamed $J_{t} \rightarrow J$)

\begin{eqnarray}
N_{i} (J,t) = & N_{0} \; A_{L}(t+1-i) \; F(J,t) , &
\mbox{if $t \ge i $} \nonumber \\
\mbox{=} & 0, & \mbox{otherwise}
\end{eqnarray}

The number of babies $N_{0}(J,t)$ may be determined using equations (3) and (9).
The mean survival probability at any time $t$ and at arbitrary age $i$ can be 
calculated as

\begin{equation}
<J>_{i} (t) = \frac{\int_{0}^{1} J \; N_{i}(J,t) \; dJ} 
{\int_{0}^{1} N_{i}(J,t) \; dJ} = \frac{\int_{0}^{1} J \; 
F(J,t) \; dJ}{\int_{0}^{1} F(J,t) \; dJ}
\end{equation} 
which is {\em independent} of $i$, that is, the model {\em does not show 
senescence !}.
It is important to note that, although ageing is absent in our model (in the 
sense that the mean survival probability does not depend on the age $i$), 
an individual can show senile decay (that is, its survival probability $J$
diminishes with time). This individual handicap is compensated
by the natural selection of the fittest so, at a collective level,
no senescence is observable. We also call attention to the fact that, 
in our model, the total number of individuals with age $i$ diminishes with $i$. 
If we integrate $N_{i}(J,t)$ over $J$ (equation (9)), the ratio 
between two successives ages $i$ and $i+1$ is given by 
$A_{L}(t+1-i)/A_{L}(t+i)$ of the Fibonacci
sequence. This ratio depends on age $i$ only for small time $t$. Moreover,
the logarithm of this ratio corresponds to the so called mortality rate.
For humans, the Gompertz law \cite{gom} suggests that mortality increases 
exponentially with age. In our model it is constant.

At a fixed age $i$ and survival probability $J$, the population increases 
with time at a rate given by
$\frac{N_{i}(J,t)} {N_{i}(J,t-1)} = \frac{A_{L}(t+1-i)}{A_{L}(t-i)} 
\cdot \frac{F(J,t)}{F(J,t-1)}$. The first ratio of the right side can be 
easily calculated by the generalized Fibonacci sequences, but the second 
requires a numerical analysis.

\section{Numerical Analysis}

\indent

The continous variable $J$ can be divided into $Q$ intervals such that 
$J \equiv j/Q $ for $j=1 \dots Q$ and infinitesimal increment $dJ \equiv 1/Q$. 
For $Q$ big enough 
we expect to reobtain the continous limit. In the same way, the products of
$f(J_{j},J_{j-1})$ in the equation (7) can be seen as simple matricial 
products. We wrote down a FORTRAN program with extended precision to determine 
$F(J,t)$ and $ < J >_{i}(t)$ at any elapsed time $t$. 
For $Q=400$, $L=10$, $a=0.04$ and $b=0.02$, Fig.1 shows the dependence of
$F(J,t)$ with $J$ at different instants. After a 
time $t > 50$ we verified that

\begin{equation}
F(J,t+1) = c \; F(J,t)
\end{equation}
where $c$ depends on $a$ and $b$ but not on $J$. This means that
after enough time, the system reaches an asymptotic limit where $ F(J,t) 
\equiv
c \; ^ {t} \bar{F}(J)$, i. e., there is a separation of variables 
and $\bar{F}(J)$ is a
stationary solution.
Fixing $b = 0.02$ and varying $a = 0.02$, $0.04$ and $0.08$, we 
determine numerically that $c = 0.92$, $0.76$ and $0.48$, respectively. If,
for example, we choose $L=10$ and $m_{i} = 1 $ for any $i$ then 
$ \lim _{t \rightarrow \infty} \frac{A_{L}(t+1)}{A_{L}(t)} = 1.9990 \dots$,
and a Malthusian exponential growth of the population $e^{r \; t}$ with
$r=0.61, 0.42$ and $-0.04$, respectively, is obtained. The last 
value shows that the system exhibits mutational meltdown \cite{ly} (extinction).

The discretized form of the variable $J$ can also be used in order to calculate 
the mean survival probability $<J(t)>$ (equation (10), dropping the now 
unnecessary lower index $i$) as a function of time. Table 1
shows our results for $L = 10$. For fixed matrix dimension, the time 
convergence is very fast.

\section{Monte Carlo Simulation and Exact Solution}

\indent

To check out all these features, we also made some simulations of the model.
Starting with an uniform distribution of the babies, we submit all of them 
to mutations as described in equation (1). For each baby, a random number 
$\epsilon$ is generated in the computer and its new survival probability $J'$ 
is calculated. Then, playing the role of the natural selection, a random number
$r$ is generated and compared with $J'$. If $r < J'$ then the baby becomes an
individual of age $1$ and produces $m_{1}$ offsprings with inherited
characteristics (that is, with the same survival probability $J'$). As the
process continues, care should be taken in order to avoid an explosion
of the computer's memory. To this end, we limited the number of 
individuals by a random decimation. Usually, in population dynamics
simulations, this decimation is interpreted as a result of food restrictions
\cite{ber}. Fig. 2 (a) shows that for $L=10$, 
$a=0.04$ and $b=0.02$, the final mean survival probability $ < J > = 
\lim_{t \rightarrow \infty} < J >(t)$ approaches $0.78$ in a very good 
agreement with our numerical results.
For $a=0.02$ and $b=0.02$ we find $ < J > \sim 0.96$.
The case $a=0.08$ (Fig.2 (c)) leads to extinction, as it was predicted.

As another important check, let us look at the Euler-Lotka equation 
\cite{rose} which, for our model, reads

\begin{equation}
\sum_{i=1}^{L} m_{i} \; \left( < J > \; e^{\; -r} \right)^{i}
\end{equation}
with $r$ being the Malthusian growth exponent.

Substituting the simulated values $<J> = 0.78$ and $<J> = 0.96$ into the
Euler-Lotka equation, we obtain
the Malthusian growth exponents $r = 0.44$ and $r = 0.65$, respectively, which
are in a fairly good agreement with those of our numerical prediction.

Incidentally, equations (11) and (7) can be consistently used to guide us to an
analytical solution of the stationary $\bar{F}(J)$ . One can write the
integral equation

\begin{equation}
c \; \bar{F}(J') = \int_{0}^{1} f(J',J) \; \bar{F}(J) \; dJ
\end{equation}

Integrating the right side and expanding the result in a Fourier series,
equation (13) turns out to be a set of linear equations for the Fourier 
coefficients. In order to have non trivial 
solutions, this set must have null determinant. The condition of zero 
determinant allow us to obtain the constant $c$. For example, if $a=0.04$
and $b=0.02$ we find $c=0.75$, in well accordance with the numerical results.
The Fourier coefficients have awful expressions so, to preserve the reader,
we will not give them. 

\section{Discussion}

\indent

Let us discuss the relevance of our model by comparing it with others
evolutionary models. The Penna model is certainly the most intensively 
investigated \cite{pen}. It exhibits ageing and sometimes catastrofic 
senescence. Contrary to what happens in our case, in the Penna model only 
babies are affected by mutations. Moreover, mutation plays the role of a
programmed death - individuals which have accumulated a number of mutations 
(i. e., number of 1's in the bit-string) larger than a threshold $T$ die.
This fate can be anticipated only if the individual dies by food restrictions
(Verhulst factor, which acts irrespective of individual fitness). In our model
there is not such a threshold and individuals may live longer. Besides that,
natural selection here operates in a very hard (and explicit) way to eliminate 
those individuals which have suffered bad mutations. Also exact solutions
have been found for the Penna model \cite{piza}. The evolution equations 
involve directly the Verhulst factor. We have similar equations ((3) and (4)), 
but with the survival probability $J$ instead. Amazingly, the Leslie matrices
were also found in the Penna model but in a different context. There, the
elements of these matrices are connected to the mutation rate while in our
model they are associated to the birth rate.

As we said in the beginning, there are
two kinds of evolutionary theories: optimal and mutational. Our model belongs
to the latter. The optimality theory is based on the fact that some genes
have antagonistics effects, that is, they can be very beneficial early in life 
but deleterious late. For example, genes enhancing early survival by promoting
a bone hardening might reduce later survival by promoting arterial hardening.
These ideas were completely embodied by the Partridge-Barton model 
\cite{pb}. 
Further studies on this model, have incorporated somatic as well as 
hereditary mutations \cite{sta,ray}, leaving the model with two mechanisms of
senescence: antagonistic pleiotropy and accumulation of bad mutations. 
But ageing (due to mutations) emerges in these works as a result of turning 
more intense with age (in some artificial or arbitrary way) the mutational 
strength. Their procedures would be equivalent to assume, in our model, that 
the $\epsilon$ (the mutational control parameter of eq. (1)) is a function 
of the age $i$. Clearly, this would also trigger an ageing process in our model.

More interesting, however, is a different version of the Partridge-Barton
model which is called Vollmar-Dasgupta \cite{vodas} model. It was 
generalized by Heumann and H\"otzel \cite{hh} to support many age intervals.
In this model, each individual carries, like its genome, a whole set of 
{\em independent} survival
probabilities $\{ J_{0}, \dots ,J_{L} \} $ where $J_{i}$ ($i=0 \dots L$) is
the {\em actual} survival probability at age $i$. 
Deleterious mutations can now affect any of them but only those coincident
with the actual age $i$ will pass through natural selection. This means that
most of the damage will only manifest later on. This accumulation of harmful 
mutations leads to senescence. Our model differs from 
Heumann and H\"otzel only in the 
point that our individuals carry just one survival probability - that of its 
actual age. This is sufficient to change radically the results.

In summary, although containing all the relevant features of evolutionary 
systems like age-structure, advantageous or deleterious mutations, 
reproduction with inherited characteristics and natural selection, our model
does not show senescence. In this way, it is a good 
candidate to a appropiately describe some coelenterates and prokaryotes 
groups, since all them appear to lack senescence. On the other hand, 
the beautiful analytical solution that we find and the techniques involved, 
encourage us to look forward new and more sophisticated models. 

\section{Acknowledgements}

\indent

We acknowledge CNPq (Conselho Nacional de Desenvolvimento Cient\'{\i}fico e 
Tecnol\'ogico) for the financial support.


\newpage 

\begin{center} 
TABLE CAPTION 
\end{center}

\vspace{2.0cm}

Table 1. The mean survival probability $<J>$, obtained by using our matricial 
formalism, as a function of the matrix dimension $Q$ and time $t$ for $L=10$, 
$a=0.04$ and $b=0.02$.

\newpage

\begin{center} 
FIGURE CAPTION 
\end{center}

\vspace{2.0cm}

Figure 1. Plots of the function $F(J,t)$ for $t=100$, $200$, $400$ and 
$800$ against
the survival probability $J$. In order to bring them to the same scale, they
were multiplied by a factor $s =  10^{11}$, $10^{23}$, $10^{47}$ and $10^{95}$,
respectively.

\vspace{2.0cm}

Figure 2. (a) time evolution of the simulated values of $<J>(t)$ for
$L=10$, $a=0.04$ and $b=0.02$. The eleven age curves coincide and are 
undistinguishable, (b) the corresponding population plotted against the time. 
The stationary behavior is an artefact of the decimation process, (c) the 
same for $L=10$, $a=0.08 $ and $b=0.02$, (d) the corresponding population
comes to extinction.

\newpage 

\begin{center} 
Table 1 
\end{center}

\begin{center}
\begin{tabular}{||c|c|c|c|c|c||} \hline\hline
\multicolumn{1}{||c|}{$ Q \; \backslash \; t $} & \multicolumn{1}{|c|}{50}
& \multicolumn{1}{|c|}{100} & \multicolumn{1}{|c|}{200} 
& \multicolumn{1}{|c|}{400} & \multicolumn{1}{|c||}{800}\\ \hline
50   &  0.539056 & 0.508864 & 0.503227 & 0.501411 & 0.500663 \\ \hline
100  &  0.692779 & 0.607288 & 0.577722 & 0.575042 & 0.575013 \\ \hline
200  &  0.751139 & 0.736371 & 0.734567 & 0.734534 & 0.734534 \\ \hline
400  &  0.771775 & 0.765403 & 0.765341 & 0.765341 & 0.765341 \\ \hline\hline
\end{tabular} 
\end{center}

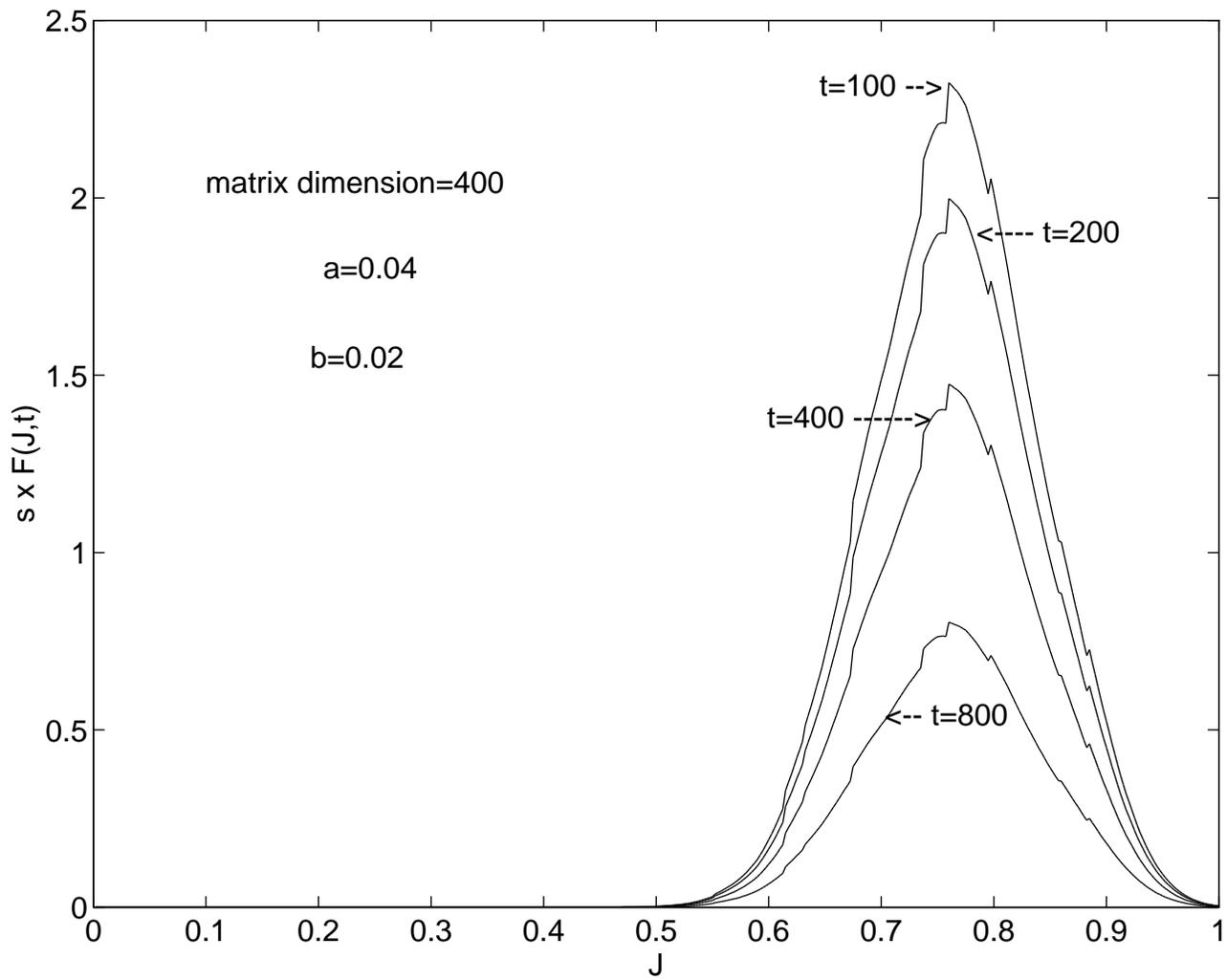
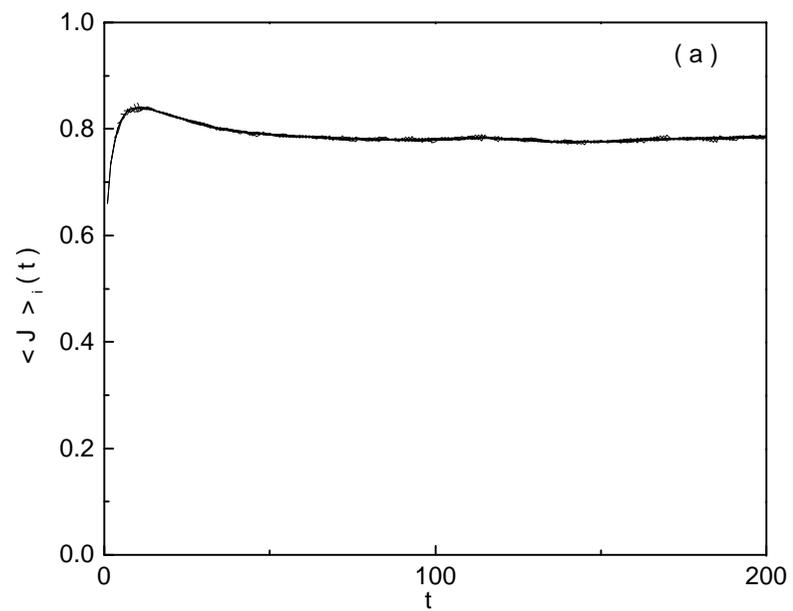
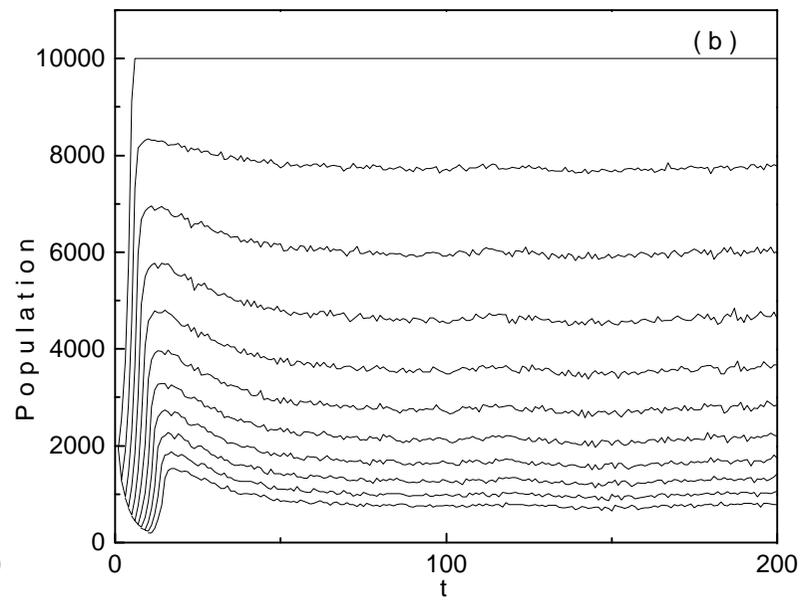
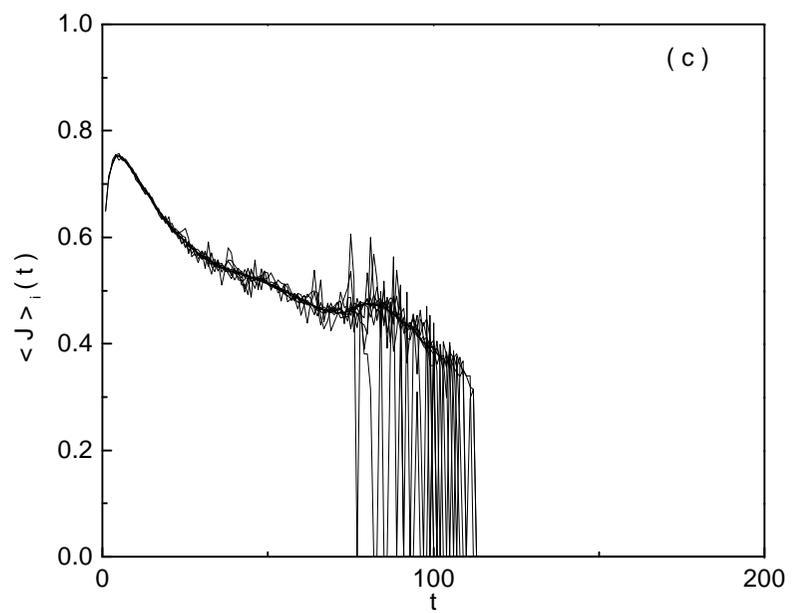
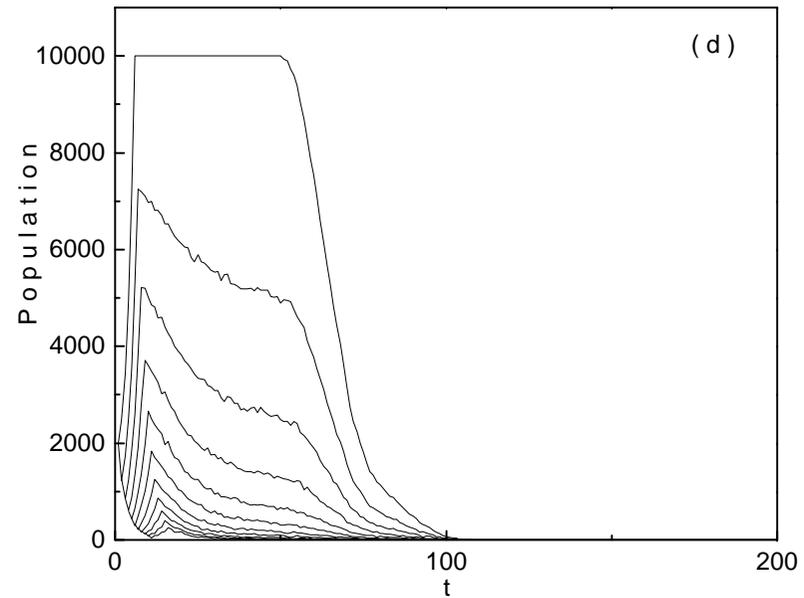
\end{document}